\documentclass[aps,pra,superscriptaddress,twocolumn,nofootinbib]{revtex4-1}
\usepackage{amsmath,bbm}
\usepackage{amssymb}
\usepackage{graphicx}
\usepackage{color}
\usepackage{amsthm}
\usepackage{ dsfont }
\usepackage{float}
\usepackage{cancel}

\usepackage[colorinlistoftodos]{todonotes}
\usepackage[colorlinks=true, allcolors=blue]{hyperref}
\usepackage{amsthm}
\usepackage{amsmath}
\usepackage{amssymb}
\usepackage{graphicx}
\usepackage{color}
\usepackage{enumitem}
\usepackage[normalem]{ulem}

\newtheorem*{theorem*}{Theorem}

\newcommand{\be}{\begin{equation}}
\newcommand{\ee}{\end{equation}}
\newcommand{\ba}{\begin{eqnarray}}
\newcommand{\ea}{\end{eqnarray}}

\newcommand{\jd}[1]{#1}

\begin{document}

\renewcommand{\today}{\number\day\space\ifcase\month\or
   January\or February\or March\or April\or May\or June\or
   July\or August\or September\or October\or November\or December\fi
   \space\number\year}

\title{Constraints on nonlocality in networks 
from \jd{no-signalling} and independence}

\date{\today}

\author{Nicolas Gisin}
\affiliation{D\'epartement de Physique Appliqu\'ee, Universit\'e de Gen\`eve, CH-1211 Gen\`eve, Switzerland}

\author{Jean-Daniel Bancal}
\affiliation{D\'epartement de Physique Appliqu\'ee, Universit\'e de Gen\`eve, CH-1211 Gen\`eve, Switzerland}

\author{Yu Cai}
\affiliation{D\'epartement de Physique Appliqu\'ee, Universit\'e de Gen\`eve, CH-1211 Gen\`eve, Switzerland}

\author{Patrick Remy}
\affiliation{D\'epartement de Physique Appliqu\'ee, Universit\'e de Gen\`eve, CH-1211 Gen\`eve, Switzerland}

\author{Armin Tavakoli}
\affiliation{D\'epartement de Physique Appliqu\'ee, Universit\'e de Gen\`eve, CH-1211 Gen\`eve, Switzerland}

\author{Emmanuel Zambrini Cruzeiro}
\affiliation{D\'epartement de Physique Appliqu\'ee, Universit\'e de Gen\`eve, CH-1211 Gen\`eve, Switzerland}
\affiliation{Laboratoire d'Information Quantique (LIQ), Université Libre de Bruxelles, 1050 Bruxelles, Belgium}

\author{Sandu Popescu}
\affiliation{H.H. Wills Physics Laboratory, Tyndall Avenue, BS8 1TL, Bristol, UK}
\affiliation{Institute for Theoretical Studies, ETH Zurich, Switzerland}

\author{Nicolas Brunner}
\affiliation{D\'epartement de Physique Appliqu\'ee, Universit\'e de Gen\`eve, CH-1211 Gen\`eve, Switzerland}

\begin{abstract}
The possibility of Bell inequality violations in quantum theory had a profound impact on our understanding of the correlations that can be shared by distant parties. Generalising the concept of Bell nonlocality to networks leads to novel forms of correlations, the characterization of which is however challenging. Here we investigate constraints on correlations in networks under the natural assumptions of no-signalling and independence of the sources. We consider the ``triangle network'' \jd{with binary outputs}, and derive strong constraints on correlations even though the parties receive no input, i.e.~each party performs a fixed measurement. We show that some of these constraints are tight, by constructing explicit local models (i.e.~where sources distribute classical variables) that can saturate them. However, we also observe that other constraints can apparently not be saturated by local models, which opens the possibility of having nonlocal (but non-signalling) correlations in the triangle network \jd{with binary outputs}.
\end{abstract}
\maketitle

 \section*{Introduction}

The no-signalling principle states that instantaneous communication at a distance is impossible. This imposes constraints on the possible correlations between distant observers. Consider the so-called Bell scenario \cite{bell}, where each party performs  different local measurements on a shared physical resource distributed by a single common source. In this case, the no-signalling principle implies that the choice of measurement (the input) of one party cannot influence the measurement statistics observed by the other parties (their outputs). In other words, the marginal probability distribution of each party (or subset of parties) must be independent of the input of any other party. These are the well-known no-signalling conditions, which represent the weakest conditions that correlations must satisfy in any ``reasonable'' physical theory \cite{PR}, in the sense of being compatible with relativity. More generally, the no-signalling principle ensures that information cannot be transmitted without any physical carrier. This provides a useful framework to investigate quantum correlations (which obviously satisfy the no-signalling conditions, but do not saturate them in general\jd{~\cite{PR}}) within a larger set of physical theories satisfying no-signalling; see e.g.~\cite{PR,barrettPR,vanDam,brassard,barrett,IC,review,sandu}.

\begin{figure}
\includegraphics[width = 0.9\columnwidth]{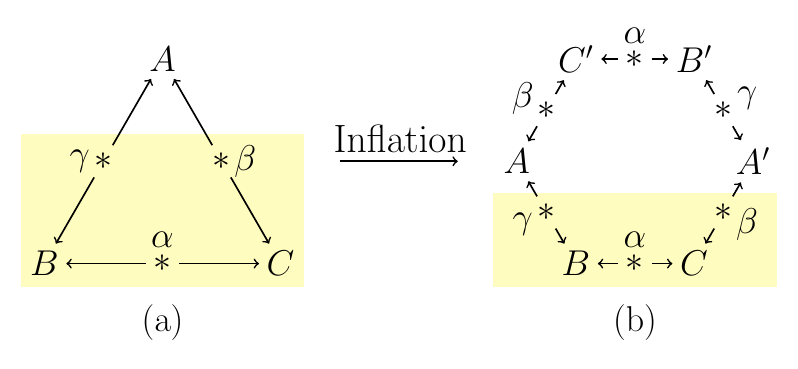}
\caption{Inflation of the triangle network to the hexagon network -- In order to capture NSI constraints in the triangle network (a), we consider an inflation to the hexagon network (b). Importantly, from the point of view of Bob and Charlie, the two situations must be indistinguishable. If not, then Alice could (instantaneously) signal to Bob and Charlie, simply by locally modifying the network structure. }
\label{fig:threeparties}
\end{figure}

Recently, the concept of Bell nonlocality has been generalized to networks, where separated sources distribute physical resources to subsets of distant parties (see Fig.~\ref{fig:threeparties}). Assuming the sources to be independent from each other \jd{\cite{branciard,fritz}}, arguably a natural assumption in this context, leads to many novel effects. Notably, it becomes now possible to demonstrate quantum nonlocality without the use of measurement inputs \cite{fritz,branciard2,gisin,wolfe2,Renou19}, but only by considering the output statistics of fixed measurements. Just recently, a first example of such nonlocality genuine to networks was proposed~\cite{Renou19,Pusey19}. This radically departs from the standard setting of Bell nonlocality, and opens \jd{many} novel questions. Characterizing correlations in networks (local or quantum) is however still very challenging at the moment, despite recent progress \cite{chaves2,armin,rosset,chaves,armin2,wolfe,lee,Denis,navascues,luo,chaves3,pozas}.

\jd{Moving beyond quantum correlations,} this naturally raises the question of finding the limits of possible correlations in networks, assuming only no-signalling and independence of the sources \cite{Lal,fritz2,Budroni,wolfe,Renou,Weilenmann}. Here we investigate this question and derive limits on correlations, which we refer to as NSI constraints (no-signalling and independence). 
\jd{While our approach can in principle be applied to any network, we focus here on the well-known triangle network with binary outputs and no inputs, for which we obtain strong, and even tight NSI constraints. Specifically, we show that, despite the absence of an input, some statistics imply the possibility for one party to signal to others by locally changing (or not changing) the structure of the network. Formally, this amounts to \jd{considering} a specific class of so-called network inflations, as introduced in Ref. \cite{wolfe}, which we show can lead to general and strong NSI constraints. Moreover, we prove that some of our NSI constraints are in fact tight, by showing that they can be saturated by correlations from explicit ``trilocal'' models, in which the sources distribute classical variables. Interestingly, however, it appears that not all of our NSI constraints can be saturated by trilocal models, which opens the possibility of having nonlocal (but nevertheless non-signalling) correlations in the triangle network with binary outputs}. Finally, we conclude with a list of open questions.


\section*{NSI constraints}

The triangle network (sketched in Fig.~\ref{fig:threeparties}(a)) features three observers: Alice, Bob and Charlie. Every pair of observers is connected by a (bipartite) source, providing a shared physical system. Importantly, the three sources are assumed to be independent from each other. Hence, the three observers share no common (i.e.~tripartite) piece of information. Based on the received physical resources, each observer provides an output ($a$, $b$ and $c$, respectively). Note that the observers receive no input in this setting, contrary to standard Bell nonlocality tests. The statistics of the experiment are thus given by the joint probability distribution $p(a,b,c)$. We focus on the case of binary outputs: $a,b,c \in \{ +1, -1\}$. It is then convenient to express the joint distribution as follows:

\ba \label{triangle}
p(a,b,c) = &\frac{1}{8} \bigg(    1 + a E_\mathrm{A} + b E_\mathrm{B} + c E_\mathrm{C} + ab E_\mathrm{AB}  \nonumber \\
& + ac E_\mathrm{AC} + bc E_\mathrm{BC} + abc E_\mathrm{ABC}   \bigg) \,,
\ea
where $E_\mathrm{A}$, $E_\mathrm{B}$ and $E_\mathrm{C}$ are the single-party marginals, $E_\mathrm{AB}$, $E_\mathrm{BC}$ and $E_\mathrm{AC}$ the two-party marginals, and $E_\mathrm{ABC}$ is the three-body correlator. Note that the positivity of $p(a,b,c)$ implies constraints on marginals, in particular $p(+++)+p(---)\geq 0$ implies
\be \label{positivity}
E_\mathrm{AB}   +  E_\mathrm{AC} +  E_\mathrm{BC} \geq -1 \,.
\ee

In the following, we will derive non-trivial constraints bounding and relating the single-party and two-party marginals of $p(a,b,c)$ under the assumption of \jd{NSI\@.} While it seems a priori astonishing that the no-signalling principle can impose constraints in a Bell scenario featuring no inputs for the parties, we will see that this is nevertheless the case in the triangle network. 

The main idea is the following. Although one party (say Alice) receives no input, she could still potentially signal to Bob and Charlie by locally modifying the structure of the network. To see this, consider the hexagon network depicted in Fig.~\ref{fig:threeparties}(b), and focus on parties Bob and Charlie. From their point of view, the two networks (triangle and hexagon) should be indistinguishable. This is because all the modification required to bring the triangle network to the hexagon (e.g.~by having Alice adding extra parties and sources) occurs on Alice's side, and can therefore be space-like separated from Bob and Charlie. If Alice, by deciding which network to use, could remotely influence the statistics of Bob and Charlie, this would clearly lead to signalling. Hence we conclude that the local statistics of Bob and Charlie (i.e.~the single-party marginals $E_\mathrm{B}$ and $E_\mathrm{C}$, as well as the two-party marginals $E_\mathrm{BC}$) must be the same in the triangle and in the hexagon. \jd{To see that this condition really captures the possibility to signal, we could imagine a thought experiment in which we would give an input to Alice, which determines whether she modifies her network structure or not. If she does so and this has an incidence on the $E_\mathrm{BC}$ marginal, then Bob and Charlie can learn about Alice's input, hence breaking the usual notion no-signalling condition. Note that the input considered here is however purely fictional, Alice's input is not present in the actual experiment.}

From the above reasoning, we conclude that the joint output probability distribution for the hexagon, i.e.~$p(a,b,c,a',b',c')$, must satisfy several constraints. In particular, one should have that

\begin{align}
\sum b \, p(a,b,c,a',b',c') &= \sum b' \, p(a,b,c,a',b',c') = E_\mathrm{B}  \\
\sum c \, p(a,b,c,a',b',c') &= \sum c' \, p(a,b,c,a',b',c') = E_\mathrm{C} \\
\sum bc \, p(a,b,c,a',b',c') &= \sum b'c' \, p(a,b,c,a',b',c') = E_\mathrm{BC} 
\end{align}
where all sums go over all outputs $a,b,c,a',b',c' $. From the independence of the sources, we obtain additional constraints, namely
\ba
\sum b b' \, p(a,b,c,a',b',c') &=& E_\mathrm{B}^2 \\
\sum c c' \, p(a,b,c,a',b',c') &=& E_\mathrm{C}^2 \\
\sum b b'c  \, p(a,b,c,a',b',c') &=& E_\mathrm{BC} E_\mathrm{B}  \\
\sum b c c'  \, p(a,b,c,a',b',c') &=& E_\mathrm{BC} E_\mathrm{C} \\
\sum bc b'c' \, p(a,b,c,a',b',c') &=& E_\mathrm{BC}^2  \,.
\ea
Clearly, we also get similar constraints when considering signalling between any other party (Bob or Charlie) to the remaining two.

Altogether, we see that NSI imposes many constraints on $p(a,b,c,a',b',c')$. Obviously, we also require that
\be  \label{norm}
p(a,b,c,a',b',c') \geq 0   \quad  \textrm{and} \quad  \sum p(a,b,c,a',b',c') = 1 \,.  
\ee
Now reversing the argument, we see that the non-negativity of $p(a,b,c,a',b',c')$ imposes non-trivial constraints relating the single- and two-party marginals of the triangle distribution $p(a,b,c)$. To illustrate this, let us proceed with an example in a slightly simplified scenario, assuming all single-party marginals to be uniformly random, i.e.~$E_\mathrm{A}=E_\mathrm{B}=E_\mathrm{C}=0$. In this case, we obtain

\begin{widetext}
\ba 
&    64\,p(a,b,c,a',b',c') = 1 + (ab+a'b')E_\mathrm{AB} + (bc+b'c')E_\mathrm{BC} + (ca'+c'a)E_\mathrm{AC} +  (abc+a'b'c')F_3 + (bca'+b'c'a)F_3'   \nonumber \\
  &       + (ca'b'+c'ab)F_3'' + aa'bb'E_\mathrm{AB}^2 + bb'cc'E_\mathrm{BC}^2 + aa'cc'E_\mathrm{AC}^2
    + aa'(bc+b'c')F_4 + bb'(ca'+c'a)F_4' + cc'(ab+a'b')F_4''   \nonumber \\
    & + aa'bb'(c+c')F_5 + bb'cc'(a+a')F_5' + aa'cc'(b+b')F_5'' +aa'bb'cc'F_6 \geq 0  \label{hexagon}
\ea
\end{widetext}
Importantly, notice that the above expression contains a number of variables (of the form $F_\mathrm{X}$) that are uncharacterized; these represent $\mathrm{X}$-party correlators in the hexagon network\jd{, see Supplementary Note 1 for more details}. Hence we obtain a set of inequalities imposing constraints on our variables of interest (i.e.~$E_\mathrm{AB}$, $E_\mathrm{BC}$ and $E_\mathrm{AC}$), but containing also additional variables which we would like to discard. This can be done systematically via the algorithm of Fourier-Motzkin elimination \cite{ziegler}. Note that here we need to treat the squared terms, such as $E_\mathrm{AB}^2$, as new variables, independent from $E_\mathrm{AB}$, so that we get a system of linear inequalities. Solving the latter, and taking into account positivity constraints as in Eq.~\eqref{positivity}, we obtain a complete characterization of the set of two-body marginals (i.e.~$E_\mathrm{AB}$, $E_\mathrm{BC}$ and $E_\mathrm{AC}$) that are compatible with NSI in the triangle network (for a hexagon inflation and uniform single-party marginals), in terms of a single inequality
\be \label{single}
 (1-E_\mathrm{AB})^2 -E_\mathrm{BC}^2 -E_\mathrm{AC}^2  \geq 0 \,,
 \ee
 and its symmetries (under relabeling of the parties and of the outputs). This implies a more symmetric, but slightly weaker inequality:
\be \label{nice}
(1+E_\mathrm{AB})^2 + (1+E_\mathrm{BC})^2 + (1+E_\mathrm{AC})^2  \leq 6 \,.
\ee
Note that when $E_\mathrm{AB}=E_\mathrm{BC}=E_\mathrm{AC} \equiv E_2$, we get simply $E_2 \leq \sqrt{2}-1 \approx 0.41$.

\begin{figure}[t!]
\includegraphics[width=\columnwidth]{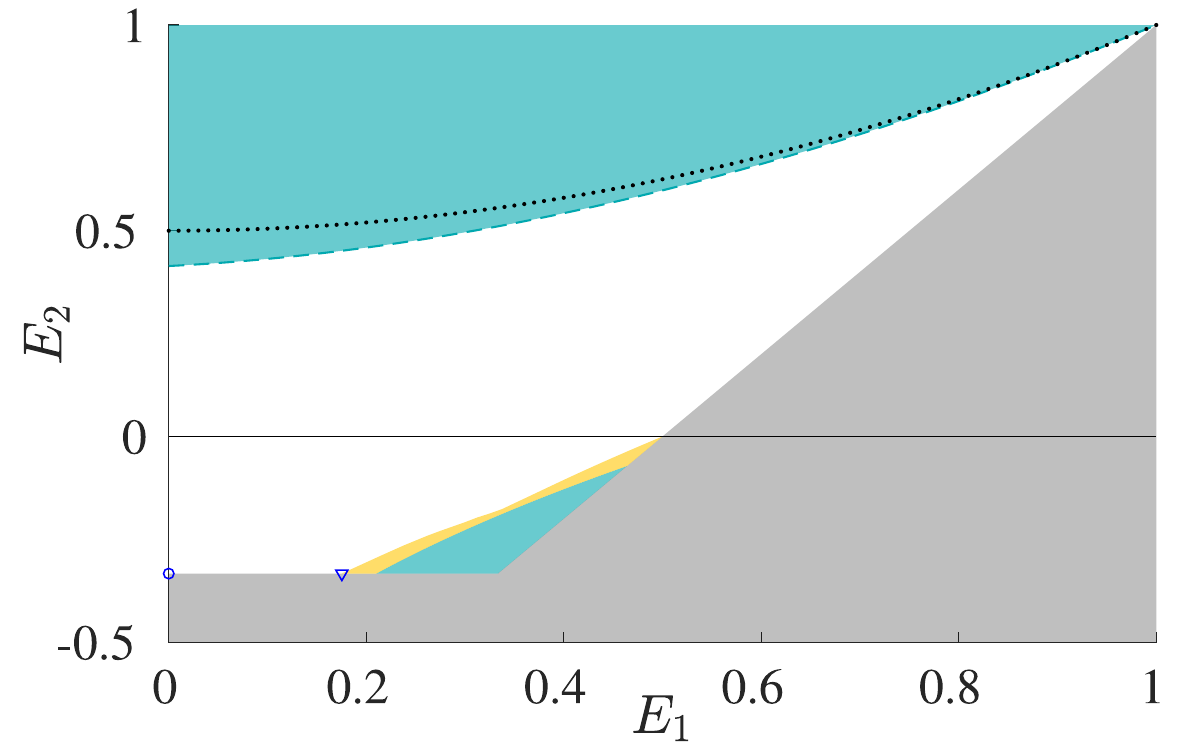}
\caption{Region of allowed correlations for symmetric distributions; projection in the plane $E_2$ vs $E_1$. The turquoise region is ruled out by NSI constraints, while the grey region is excluded from simple positivity constraints. The white region is accessible via trilocal models. Correlations in the yellow region satisfy NSI constraints (from the hexagon inflation), but we could not find a trilocal model for them. \jd{The constraint Eq.~(34) of~\cite{wolfe} is shown in dotted black.} The \jd{dashed} turquoise curve corresponds to the NSI inequality \eqref{nice2}, which turns out to be tight. Explicit trilocal models are also obtained for the correlations marked by blue dots (see Supplementary Note 2).}
\label{fig:E2vsE1}
\end{figure}

Next we consider the symmetric case (i.e.~$E_\mathrm{A}=E_\mathrm{B}=E_\mathrm{C} \equiv E_1$ and $E_\mathrm{AB}=E_\mathrm{BC}=E_\mathrm{AC} \equiv E_2$) and obtain non-trivial NSI constraints on the possible values of $E_1$ and $E_2$ (see Fig.~\ref{fig:E2vsE1}). In particular, correlations compatible with NSI must satisfy the following inequality
\be \label{nice2}
(1+ 2 |E_1| + E_2)^2 \leq 2 (1+|E_1|)^3 \,.
\ee

Let us move now to the most general case, with arbitrary values for single- and two-party marginals. For a given set of
values $E_{A}$, $E_{B}$, $E_{C}$, $E_\mathrm{AB}$, $E_\mathrm{BC}$ and $E_\mathrm{AC}$, it is possible here to determine via a linear program whether this set is compatible with NSI or not (see Supplementary Note 1). 
More generally, obtaining a characterization of the NSI constraints in terms of explicit inequalities (as above) is challenging, due mainly to the number of parameters and nonlinear constraints. We nevertheless \jd{obtain that} the following inequality represents an NSI constraint
\jd{ \be \label{conjecture}
 \begin{split}
&(1+|E_\mathrm{A}|+|E_\mathrm{B}|+E_\mathrm{AB})^2 \\
&+ (1+|E_\mathrm{A}|+|E_\mathrm{C}|+E_\mathrm{AC})^2  \\
&+ (1+|E_\mathrm{B}|+|E_\mathrm{C}|+E_\mathrm{BC})^2 \\
&\leq 6(1+|E_\mathrm{A}|)(1+|E_\mathrm{B}|)(1+|E_\mathrm{C}|) \,.
\end{split}
\ee}
\jd{A proof of this general inequality is given in Supplementary Note 4.} Note that this inequality reduces to \eqref{nice} when $E_\mathrm{A}=E_\mathrm{B}=E_\mathrm{C}=0$, as well as to \eqref{nice2} for the symmetric case.

It is worthwhile discussing the connection between our approach and the ``inflation technique'' presented in Refs.~\cite{wolfe,navascues}. There, the main focus is on using inflated networks for deriving constraints on correlations achievable with classical resources. In \jd{that} case, information can be readily copied, so that sources can send the same information to several parties. Ultimately, this allows for a full characterization of correlations achievable with classical resources \cite{wolfe}. Copying information is however not possible in our case, as no-signalling resources cannot be perfectly cloned in general \cite{barrett}. Hence only inflated networks with bipartite sources can be considered in our case, such as the hexagon. A discussion of these ideas can be found in Section V.D of \cite{wolfe}, where the idea of using inflation to limit no-signalling correlations in networks is mentioned. Here, we derive explicitly bounds that all correlations satisfying the NSI constraints, whether quantum of post-quantum, have to satisfy\jd{, and identify the physical principle behind them}.

Finally, the choice of the hexagon inflation deserves a few words. As seen from Fig.~\ref{fig:threeparties}(b), it is judicious to consider inflated networks forming a ring, with a number of parties that is a multiple of three. Intuitively this should enforce the strongest constraints on the correlations of the inflated network; in particular, all single and two-body marginals are fixed by the correlations of the triangle. This would not be the case when considering inflations to ring networks with a \jd{number of parties that is not divisible by three.}

\section*{Tightness}
A natural question is whether the constraints we derived above, that are necessary to satisfy NSI, are also sufficient. There is a priori no reason why this should be the case. Of course, starting from the triangle network, there are many (in fact infinitely many) possible extended networks that can be considered, and no-signalling must be enforced in all cases. For instance, instead of extending the network to a hexagon (as in Fig.~\ref{fig:threeparties}), Alice could consider an extension to a ring network featuring 9, 12 or more parties. Clearly, such extensions could lead to stronger constraints than those derived here for the hexagon network.

Nevertheless, we show that some of the constraints we obtain above are in fact tight, i.e.~necessary and sufficient for \jd{NSI\@.}
We prove this by presenting explicit correlations (constructed within a \jd{generalized probabilitic theory} satisfying NSI) that saturate these constraints. In fact, we consider simply the case where all sources distribute classical variables to each party, which we refer to as ``trilocal'' models. The latter give rise to correlations of the form
\begin{align} \label{trilocal}
p(a,b,c) = &\int \mu(\alpha) d\alpha \int \nu(\beta)  d\beta \int \omega(\gamma)  d \gamma   \\ & p_A(a|\beta, \gamma) \, p_B(b|\alpha, \gamma) \, p_C(c|\alpha, \beta)   \nonumber
\end{align}
where $\alpha$, $\beta$ and $\gamma$ represent the three local variables distributed by each source, with arbitrary probability densities $\mu(\alpha)$, $\nu(\beta)$ and $\omega(\gamma)$. Also, $p_A(a|\beta, \gamma)$ represents an arbitrary response function for Alice, and similarly for $p_B(b|\alpha, \gamma)$ and $p_C(c|\alpha, \beta)$. Note that such trilocal models represents a natural extension of the concept of Bell locality to networks (see e.g.~\cite{branciard,rosset}).

We first consider the case of symmetric distributions, i.e.~characterized by the two parameters $E_1$ and $E_2$, and seek to determine the set of correlations that can be achieved with trilocal models. As shown in Fig.~\ref{fig:E2vsE1}, it turns out that almost all NSI constraints can be saturated in this case, in particular the inequality \eqref{nice2}. After performing a numerical search, we could construct explicitly some of these trilocal models, which involve up to ternary local variables (see Supplementary Note 2 for details). \jd{Moreover, we compare our NSI constraint \eqref{nice2} to the one derived in Ref. \cite{wolfe} (see Eq. (34)), and find that the present one is stronger, and in fact tight (see Fig.~\ref{fig:E2vsE1}). Note also that a previous work derived \jd{an} NSI constraint based on entropic quantities \cite{Lal}; such constraints are however known to be generally weak, as entropies are a coarse-graining of the statistics, which no longer distinguishes between correlations and anti-correlations.}

As seen from Fig.~\ref{fig:E2vsE1}, there is however a small region (in yellow) which is compatible with NSI (considering the hexagon inflation), but for which we could not construct a trilocal model. Whether this gap can be closed by considering more sophisticated local models (using variables of larger alphabet) or whether stronger no-signalling bounds can be obtained is an interesting open question. For the triangle network with binary outcomes, any trilocal distribution can be obtained by considering shared variables of dimension (at most) six, and deterministic response functions \cite{Denis}.

\begin{figure}[t!]
\includegraphics[width = \columnwidth]{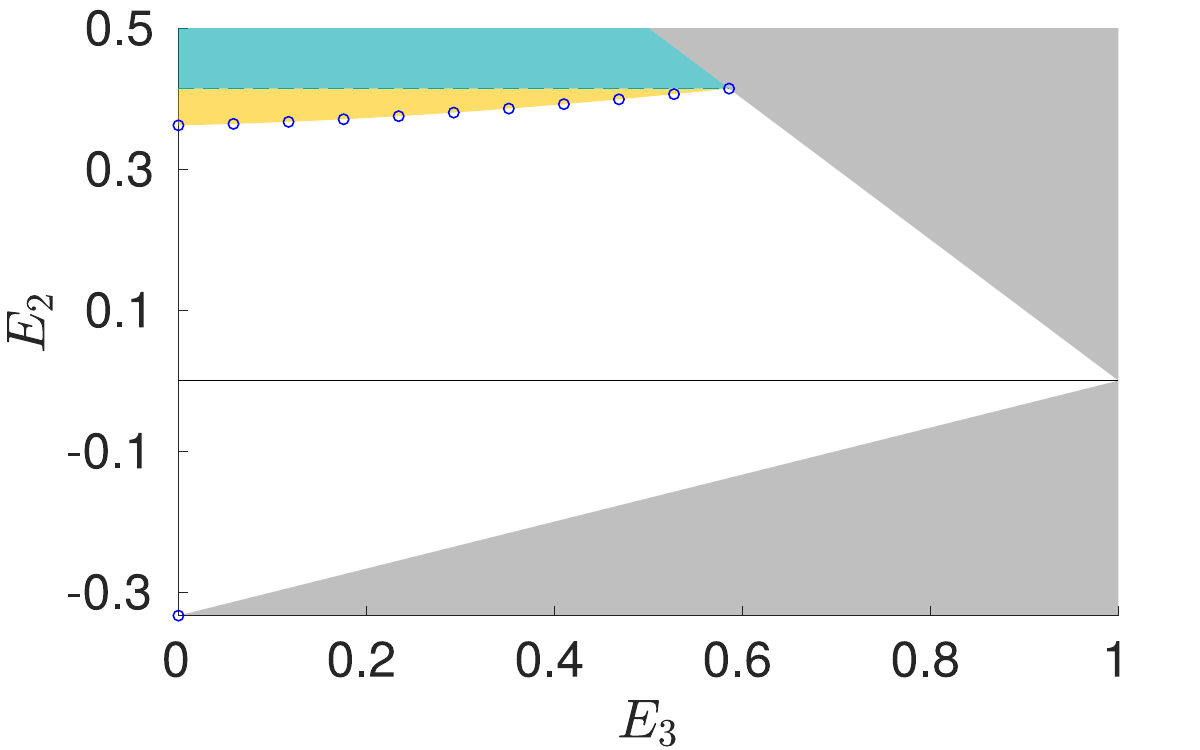}
\caption{Region of allowed correlations for symmetric distributions with $E_1=0$; represented in the plane $E_2$ vs $E_3$. The turquoise region is ruled out by NSI constraints (\jd{dashed} turquoise \jd{line} given by Eq.~\eqref{nice2}), while the grey region is excluded from simple positivity constraints. The white region is accessible via trilocal models. Correlations in the yellow region satisfy NSI constraints (from the hexagon inflation), but we could not find a trilocal model for them. Explicit trilocal models are also obtained for the correlations marked by blue dots (see Supplementary Note~2).}
\label{fig:E2vsE3}
\end{figure}

In fact, another (and arguably much more interesting) possibility would be that this gap cannot be closed, as it would feature correlations \jd{with binary outcomes} satisfying NSI but that are nevertheless non-trilocal. To further explore this question, let us now focus on the case where single-party marginals vanish, i.e.~$E_1 =0$. We investigate the relation between two-party marginals $E_2$ and the three-party correlator $E_3 = E_\mathrm{ABC}$, comparing NSI constraints and trilocal models. Notice that the NSI constraints \jd{we obtain here} do not involve $E_3$ (as the latter cannot be recovered within the analysis of the hexagon). Hence NSI imposes only $E_2 \leq \sqrt{2}-1$, while positivity of $p(a,b,c)$ imposes other constraints. This is shown in Fig.~\ref{fig:E2vsE3}, where we also seek to characterize the set of correlations achievable via trilocal models (proceeding as above). Interestingly, we find again a potential gap between trilocal correlations and NSI constraints. This should however be considered with care. First, the NSI constraints obtained from the hexagon may not be optimal (see discussion). Second, there could exist more sophisticated trilocal models (e.g.~involving higher dimensional variables) that could lead to a stronger correlations (i.e.~cover a larger region in Fig.~\ref{fig:E2vsE3}). Note also that we investigated whether quantum distributions satisfying the independence assumption exist outside of the trilocal region, but we could not find any example (we performed a numerical search, considering entangled states of dimension up to $4 \times 4$).

Finally, note that we also performed a similar analysis for the case where single-party marginals vanish but two-body marginals are not assumed to be identical to each other. Here we find that inequality \eqref{single} can be saturated in \jd{a} few specific cases. However, there also exist correlations satisfying the NSI bounds that do not seem to admit a trilocal model; details in Supplementary Note~1.

\section*{Discussion}

We discussed the constraints arising on correlations in networks, under the assumption of no-signalling and independence of the sources. We focused our attention on the triangle network with binary outputs for which we derived strong  constraints, including tight ones. Our work raises a number of open questions that we now discuss further.

A first question is whether the constraints we derive (necessary under NSI), could also be sufficient. We believe this not to be the case, as stronger NSI constraints could arise from inflations of the triangle to more complex networks (e.g.~loop networks with an arbitrary number of parties). Note that there could also exist different forms of no-signalling constraints, that cannot be enforced via inflation. In this respect, we compare in Supplementary Note 3 our NSI constraints with the recent work of Ref.~\cite{Renou} proposing a very different approach to this problem, using the the Finner inequality. A notable difference is that the latter imposes constraints on tripartite correlations, which is not the case here.

Another important question is whether there could exist nonlocality in the simplest triangle network with binary outcomes. That is, can we find a $p(a,b,c)$ that satisfies NSI but that is nevertheless non-trilocal?
While we identified certain potential candidate distributions for this, we could not prove any conclusive result at this point. 
We cannot exclude the possibilities that (i) these correlations are in fact not compatible with NSI (as there exist stronger NSI constraints) or (ii) these correlations can in fact be reproduced by a trilocal model. In order to address point (i), one could try to reproduce these correlations via an explicit NSI model, for instance considering that all sources emit no-signalling resources (such as nonlocal boxes \cite{PR}) which could then be ``wired together'' by the parties. To address point (ii), one could show that these correlations violate a multilocality inequality for the triangle network. Of course finding such inequalities is notably challenging, see e.g.~\cite{gisin}.

Furthermore, it would be interesting to derive NSI constraints for other types of networks. Indeed, the approach developed here can be straightforwardly used. Cases of high interest are general loop networks, as well as the triangle
network with larger output alphabet (where examples of quantum nonlocality are proven to exist \cite{fritz,Renou19}).

Finally, a more fundamental question is whether any correlation satisfying the complete NSI constraints can be realized within an explicit physical theory satisfying no-signalling (the latter are usually referred to as generalized probabilistic theories \cite{barrett}). While this is the case in the standard Bell scenario (where all parties share a common resource), it is not clear if that would also be the case in the network scenario.

\section*{Acknowledgements}
We thank Stefano Pironio, Marc-Olivier Renou, Denis Rosset and Elie Wolfe for discussions. We acknowledge financial support from the Swiss national science foundation (Starting grant DIAQ, NCCR-QSIT and NCCR-Swissmap). E. Zambrini acknowledges support by the Swiss National Science Foundation via the Mobility Fellowship P2GEP2\_188276.

\section*{Appendix 1: NSI bounds}

In the hexagon configuration introduced in the main text, the probability distribution can be expressed in terms of 16 parameters as
\begin{equation}\label{eq:hexagonDecomp}
\begin{split}
6&4\,p(a,b,c,a',b',c') = \\
&\ 1 + (a+a')E_\mathrm{A} + (b+b')E_\mathrm{B} + (c+c')E_\mathrm{C}\\
&+ (ab'+a'b)E_\mathrm{A} E_\mathrm{B} + (bc'+b'c)E_\mathrm{B} E_\mathrm{C} + (ac+a'c')E_\mathrm{A} E_\mathrm{C}\\
&+ aa'E_\mathrm{A}^2 + bb'E_\mathrm{B}^2 + cc'E_\mathrm{C}^2 + (ab'c + a'bc')E_\mathrm{A} E_\mathrm{B} E_\mathrm{C}\\
&+ (ab+a'b')E_\mathrm{AB} + (bc+b'c')E_\mathrm{BC} + (ca'+c'a)E_\mathrm{AC}\\
&+ aa'(b+b')E_\mathrm{A} E_\mathrm{AB} + aa'(c+c')E_\mathrm{A} E_\mathrm{AC}\\
&+ bb'(a+a')E_\mathrm{B} E_\mathrm{AB} + bb'(c+c')E_\mathrm{B} E_\mathrm{BC}\\
&+ cc'(a+a')E_\mathrm{C} E_\mathrm{AC} + cc'(b+b')E_\mathrm{C} E_\mathrm{BC}\\
&+ aa'bb'E_\mathrm{AB}^2 + bb'cc'E_\mathrm{BC}^2 + aa'cc'E_\mathrm{AC}^2\\
&+ aa'(cb'+bc')E_\mathrm{A} F_3'' + bb'(ac+a'c')E_\mathrm{B} F_3 \\
& + cc'(ba'+b'a)E_\mathrm{C} F_3'\\
&+ (abc+a'b'c')F_3 + (bca'+b'c'a)F_3' + (ca'b'+c'ab)F_3''\\
&+ aa'(bc+b'c')F_4 + bb'(ca'+c'a)F_4' + cc'(ab+a'b')F_4''\\
&+ aa'bb'(c+c')F_5 + bb'cc'(a+a')F_5' + aa'cc'(b+b')F_5''\\
&+aa'bb'cc'F_6.
\end{split}
\end{equation}
\jd{Six} of these parameters, which are part of the \jd{behaviour} vector $\mathcal{E}=(E_\mathrm{A}, E_\mathrm{B}, E_\mathrm{C}, E_\mathrm{AB}, E_\mathrm{BC}, E_\mathrm{AC}, E_\mathrm{ABC})$, appear in the triangle as well. We refer to them as the physical parameters. The remaining 10, from $\mathcal{F}=(F_3, F_3', F_3'', F_4, F_4', F_4'', F_5, F_5', F_5'', F_6)$, are new to the hexagon. We refer to them as free variables.

\jd{In this decomposition the $E$ terms correspond to correlators that appear in the triangle scenario, c.f. Fig.~1(a) of the main text, whereas the free variables $F_X$ refer to $X$-partite correlators in the hexagon, c.f. Fig. 1(b) of the main text. For instance, in the case of tripartite correlators (i.e. with $X=3$), the hexagon network contains three distinct tripartite correlators, which we simply refer to as $F_3$, $F_3'$ and $F_3''$. The first one, $F_3$ is defined as the correlator for parties A, B, and C in the hexagon, i.e. $F_3=\sum_{a,b,c,a',b',c'} abc p(a,b,c,a',b',c')$. Since these parties are identical to $\textrm{A}'$, $\textrm{B}'$, $\textrm{C}'$,  we also have $F_3=\sum_{a,b,c,a',b',c'} a'b'c' p(a,b,c,a',b',c')$. We \jd{notice} however that this term is not identical to the tripartite correlation term $E_\mathrm{ABC}=\sum_{a,b,c} abc p(a,b,c)$ that can be measured in the triangle configuration (c.f. Fig 1(a) of the main text). Indeed, in the triangle configuration, parties A, B and C are connected by three sources $\alpha$, $\beta$, $\gamma$, but in the hexagon configuration, these three parties are only connected by the two sources $\alpha$ and $\gamma$. We thus have $F_3\neq E_\mathrm{ABC}$. Similarly, $F_3'=\sum_{a,b,c,a',b',c'} a'bc p(a,b,c,a',b',c')$ is in general different than $F_3$, because this time the three parties $\textrm{A}'$, B, C are connected by two different sources: $\alpha$ and $\beta$. For the same reason we have a third tripartite correlator $F_3''=\sum_{a,b,c,a',b',c'} a'b'c p(a,b,c,a',b',c')$ in the hexagon network, and similarly for the remaining free variables.}

The probabilities in the triangle configuration can be written in terms of the same variables, with an additional tripartite term $E_\mathrm{ABC}$, as
\begin{align}\label{eq:triangleDecomp}
8\, p(a,b,c) = &1 + a E_\mathrm{A} + b E_\mathrm{B} + c E_\mathrm{C} + ab E_\mathrm{AB}\\
& + bc E_\mathrm{BC} + ac E_\mathrm{AC} + abc E_\mathrm{ABC}\nonumber
\end{align}
In this appendix, we describe the constraints that the positivity conditions $p(a,b,c,a',b',c')\geq 0$ and $p(a,b,c)\geq 0$ imply on the first six parameters.

A first general observation is that Supplementary Equation \eqref{eq:triangleDecomp} is linear in its variables, but Supplementary Equation \eqref{eq:hexagonDecomp} is nonlinear. However, since this last expression involves no product of free variables, all nonlinearities vanish when the parameters in $\mathcal{E}$ are fixed. It is therefore always possible to test whether a behaviour given by some variables $\mathcal{E}$ is compatible with a hexagon configuration by linear programming. Concretely, this is achieved by solving the following linear program:
\begin{align}\label{eq:linprog}
\underset{\mathcal{F}}{\text{max}}\ \  & 1 &&\\
\text{s.t.}\ \  & f_6(a,b,c,a',b',c',\mathcal{E},\mathcal{F}) \geq 0\ &\forall& a,b,c,a',b',c'=\pm 1\nonumber\\
& f_3(a,b,c,\mathcal{E}) \geq 0\ &\forall& a,b,c=\pm 1,\nonumber
\end{align}
where $f_6(a,b,c,a',b',c',\mathcal{E},\mathcal{F})$ is the expression given by Supplementary Equation \eqref{eq:hexagonDecomp} and $f_3(a,b,c,\mathcal{E})$ the one given by Supplementary Equation \eqref{eq:triangleDecomp}. \jd{Note that this linear program involves a constant objective function because we are not trying to maximize any particular quantity, but we are rather interested in knowing whether the set of constraints admit a joint solution.} If this linear program is feasible, then the behaviour given by the vector $\mathcal{E}$ is compatible with the considered constraints. Otherwise, it is not. Note that this formulation in terms of linear programming would not be possible in inflations of the triangle involving eight or more parties since in this case products of the unknown variables, such as $F_3^2$, appear.

In general, we are interested in more than only testing whether a behaviour is compatible with the no-signalling constraints. In particular, we would like to find NSI inequalities. For this, we start by considering simplified situations.

\subsection*{\jd{Uniformly} random single-party marginals}
Let us consider the situation in which the outputs produced by all parties are uniformly random, i.e.~$E_\mathrm{A}=E_\mathrm{B}=E_\mathrm{C}=0$. In this case, the expression for the probabilities~\eqref{eq:hexagonDecomp} simplifies significantly. The number of free variables is unchanged, but all products of free variables with a physical parameter vanish. This allows us to understand the positivity constraints $p(abca'b'c')\geq 0$ as a set of linear constraints relating powers of physical parameters with some unknown variables. 
Inequalities involving only physical parameters can thus be obtained from this set of constraints by judiciously adding several probabilities to each other. For instance, we can write
\begin{equation}\label{eq:sum4}
\begin{split}
& 16(p(1, 1, 1, -1, -1, 1) + p(-1, -1, -1, 1, 1, -1)\\
&\ \  + p(1, 1, -1, 1, 1, 1) + p(-1, -1, 1, -1, -1, -1))\\
&= (1 + 2 E_\mathrm{AB} + E_\mathrm{AB}^2 - E_\mathrm{BC}^2 - E_\mathrm{AC}^2 + 2 F_4'' + F_6\\
&\ \ + 1 + 2 E_\mathrm{AB} + E_\mathrm{AB}^2 - E_\mathrm{BC}^2 - E_\mathrm{AC}^2 - 2 F_4'' - F_6)/2\\
&= 1 + 2 E_\mathrm{AB} + E_\mathrm{AB}^2 - E_\mathrm{BC}^2 - E_\mathrm{AC}^2 \geq 0,
\end{split}
\end{equation}
and obtain an inequality involving no free variable in $\mathcal{F}$. More generally, all constraints on the physical variables can be obtained by performing a Fourier-Motzkin elimination on the free variables.

Performing this elimination produces 24 inequalities. After taking into account the redundancy implied by the symmetries of the Bell scenario~\cite{Rosset14}, we recover the constraint~\eqref{eq:sum4}, implying that it is tight, together with two more constraints. Altogether, these form the three following families of constraints:
\begin{align}
(1 + E_\mathrm{AB})^2 - E_\mathrm{BC}^2 - E_\mathrm{AC}^2 &\geq 0 \label{eq:ineq1}\\
(1 + E_\mathrm{AB})^2 + E_\mathrm{BC}^2 + E_\mathrm{AC}^2 &\geq 0 \label{eq:ineq2}\\
1 + E_\mathrm{AB} + E_\mathrm{BC} + E_\mathrm{AC}^2 &\geq 0. \label{eq:ineq3}
\end{align}
The second inequality is a sum of squares, and therefore always true. Interestingly, the third inequality is a consequence of the first one:
\begin{equation}
\begin{split}
1 + &E_\mathrm{AB} + E_\mathrm{BC} + E_\mathrm{AC}^2\\
&\geq 1 + E_\mathrm{AB} + E_\mathrm{BC} - E_\mathrm{AC}^2\\
&= (1 +2E_\mathrm{AB} + E_\mathrm{AB}^2 - E_\mathrm{BC}^2 - E_\mathrm{AC}^2\\
&\ \ + 1 +2E_\mathrm{BC} - E_\mathrm{AB}^2 + E_\mathrm{BC}^2 - E_\mathrm{AC}^2)/2\\
&\geq 0.
\end{split}
\end{equation}
Therefore, the constraints due to the hexagon inflation boil down to a unique inequality : Supplementary Equation \eqref{eq:ineq1}.

In Supplementary Figure \ref{fig:EabvsEbc}, we plot the constraint imposed by this inequality in the $E_\mathrm{AB}$-$E_\mathrm{BC}$ plane for fixed values of $E_\mathrm{AC}$. Remarkably, most of the NSI region can be achieved by trilocal models, leaving only a small gap region (yellow area).

\begin{figure}
\centering
\includegraphics[width=0.5\textwidth]{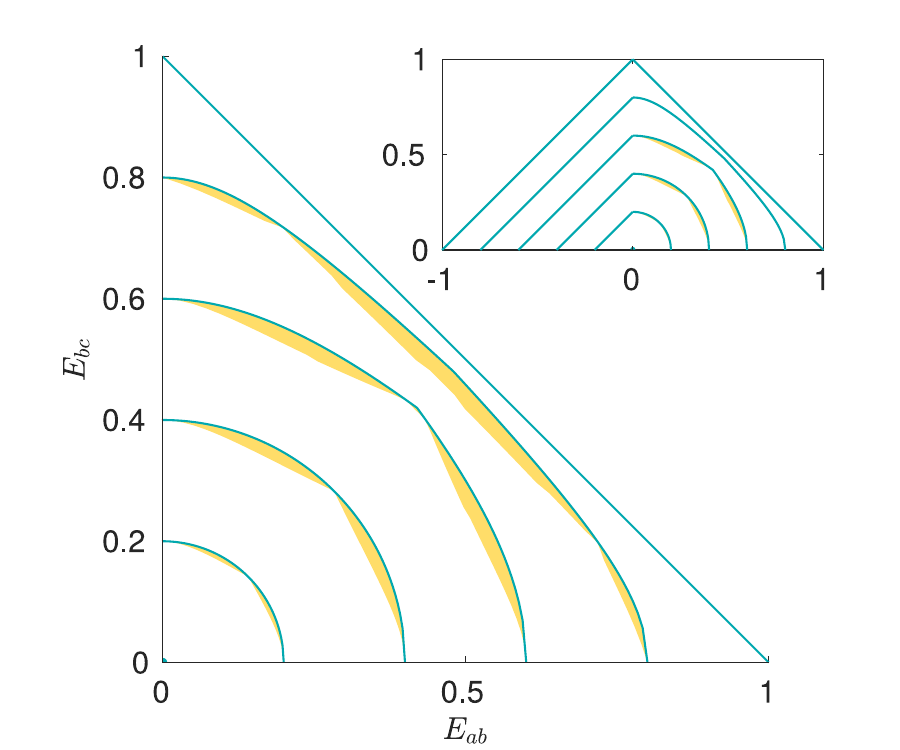}
\caption{Turquoise lines show the border of the NSI constraints imposed by Supplementary Inequality~\eqref{eq:ineq1} and the positivity $p(abc)\geq 0$ for values of $E_\mathrm{AC}$ equal to $\{0, 0.2, 0.4, 0.6, 0.8, 1\}$ (starting from the outside). When $E_\mathrm{AC}=1$, the allowed region is a single point at the origin. The inset shows the whole range of $E_\mathrm{AB}\in[-1,1]$: straight lines on the left part are positivity constraints, whereas lines on the right part correspond to Supplementary Inequality~\eqref{eq:ineq1}. The curves for $E_\mathrm{BC}\leq0$ can be obtained by letting Bob flip his output, hence the full figure is symmetric under a $\pi$ rotation around the origin. For fixed values of $E_\mathrm{AC}$, the yellow area shows the ``mystery'' region for which we could not find a trilocal model. Note that local models can reach any part of the NSI boundary for $E_\mathrm{AB}\leq 0$ and $E_\mathrm{BC}\geq 0$.}
\label{fig:EabvsEbc}
\end{figure}

Combining three versions of Supplementary Inequality~\eqref{eq:ineq1}, we obtain
\begin{equation}
\begin{split}
& 1 -2E_\mathrm{AB} + E_\mathrm{AB}^2 - E_\mathrm{BC}^2 - E_\mathrm{AC}^2\\
+ &1 -2E_\mathrm{BC} - E_\mathrm{AB}^2 + E_\mathrm{BC}^2 - E_\mathrm{AC}^2\\
+ &1 -2E_\mathrm{AC} - E_\mathrm{AB}^2 - E_\mathrm{BC}^2 + E_\mathrm{AC}^2 \geq 0
\end{split}
\end{equation}
which simplifies to the symmetric inequality
\begin{equation}
(1+E_\mathrm{AB})^2+(1+E_\mathrm{BC})^2+(1+E_\mathrm{AC})^2 \leq 6.
\end{equation}
This inequality does not detect the point $E_\mathrm{AB}=1/2$, $E_\mathrm{BC}=-3/5$, $E_\mathrm{AC}=0$, which violates Supplementary Inequality~\eqref{eq:ineq1}. The latter inequality is thus tighter.

Note that in the case where $E_\mathrm{AB}=E_\mathrm{BC}=E_\mathrm{AC}=E_2$, Supplementary Inequality~\eqref{eq:ineq1} implies $E_2\leq\sqrt{2}-1$ whereas Supplementary Inequalities~\eqref{eq:ineq2}-\eqref{eq:ineq3} are always satisfied.

We note also that the positivity constraints on the triangle impose the following condition on the bipartite marginals in presence of random marginals:
\begin{equation}\label{eq:posTriangle}
E_\mathrm{AB}+E_\mathrm{BC}+E_\mathrm{AC}\geq -1.
\end{equation}
Since the point $E_\mathrm{AB}=E_\mathrm{BC}=E_\mathrm{AC}=-\sqrt{2}+1$ satisfies inequalities in the family of Supplementary Equation ~\eqref{eq:ineq1}, but violates Supplementary Equation \eqref{eq:posTriangle}, it is not a consequence thereof.

\subsection*{Symmetric statistics}
We now consider the special case in which the single party marginals are not all zero, but the statistics are invariant under exchange of the parties. The bipartite statistics can then be parametrized by two numbers $E_1=E_\mathrm{A}=E_\mathrm{B}=E_\mathrm{C}$ and $E_2=E_\mathrm{AB}=E_\mathrm{BC}=E_\mathrm{AC}$.

Under this assumption, all free parameters are still present in the decomposition~\eqref{eq:hexagonDecomp}, but the physical space is only of dimension 2. This time, however, the probabilities involve products of known with unknown variables, like $E_1F_3$. Therefore, we cannot resort only to Fourier-Motzkin elimination to obtain all constraints that apply to the physical terms. Doing so by considering $E_1F_3$ as a free variable to eliminate indeed would not take into account the actual value of $E_1$. In particular, if $E_1=0$, this terms would already be eliminated.

We can however resort to the linear programming formulation described earlier in Supplementary Equation~\eqref{eq:linprog} to describe the set of correlations in the two-dimensional space of $E_1$-$E_2$ which are compatible with the considered constraints. This gives rise to Fig.~2 presented in the main text. In particular, we verify up to numerical precision that the upper bound on $E_2$ as a function of $E_1$ of the form
\begin{equation} \label{SM:nice2}
(1 + 2E_1 + E_2)^2 \leq 2(1+E_1)^3.
\end{equation}
As described in Supplementary Note 2, this bound is achievable.

\section*{Appendix 2: Construction of trilocal models}
\label{Sec:localModels}

In this appendix we present the explicit construction of some of the trilocal models. A trilocal model consists of (i) three distributions, for each of the shared classical variables: $\mu(\alpha)$, $\nu(\beta)$ and $\omega(\gamma)$, and (ii) three output functions (one for each party): $p_A(a|\beta, \gamma)$, $p_B(b|\alpha, \gamma)$ and $p_C(c|\alpha\beta)$. The resulting statistics is given by 
\begin{align} 
p(a,b,c) = &\int \mu(\alpha) d\alpha \int \nu(\beta)  d\beta \int \omega(\gamma)  d \gamma   \\ & p_A(a|\beta, \gamma) \, p_B(b|\alpha, \gamma) \, p_C(c|\alpha, \beta)  . \nonumber
\end{align}

In a two-outcome scenario, it is sufficient to specify the output functions for outcome 1, since $p_A(a=-1|\beta, \gamma) = 1- p_A(a=1|\beta, \gamma)$. For ease of notation we write $p_A(a=1|\beta, \gamma) \equiv f_a(\beta,\gamma)$. Without loss of generality, we may assume the same alphabet size, $d$, for all the shared variables. Hence a $d$it model can be represented by a $3$-by-$d$ matrix (with $3(d-1)$ variables) for the distributions and three $d$-by-$d$ matrices, $f_a(\beta,\gamma)$, $f_b(\alpha,\gamma)$ and $f_c(\beta,\alpha)$  (each with $d^2$ variables) for the output functions. For example,

\begin{align}
	P = \left[
	\begin{array}{c|cc}
		& 1 & 2 \\
		\hline
		\alpha & 1/2 & 1/2 \\
		\beta & 1/2 & 1/2 \\
		\gamma & 1/2 & 1/2
	\end{array} \right], \;
	f_a = f_b = f_c =
	\begin{bmatrix}
		1 & 0 \\
		0 & 1
	\end{bmatrix} \,,
\end{align}
denotes the strategies in which each source emits one bit at random and the parties output $1$ whenever they receive the same bit from the two sources they are connected to.

Given a specific distribution $p(abc)$, or a specific set of marginals (e.g.~$E_1$ and $E_2$), it is possible to numerically search for a trilocal model reproducing this data. Notably, for binary outcomes, it is sufficient to consider trilocal models with shared variables that have dimension $d\leq 6$ \cite{Denis}. We implemented this numerical procedure, and found that for the case of low dimensions (i.e.~$d=2,3$) the method is effective and appears to be reliable. For $d>3$, the method can still be run, but is less reliable (i.e.~the fact that the algorithm is not able to find a trilocal model does not necessarily mean that there exist none). We used this method to determine the trilocal regions in Supplementary Figs. 2 and 3 of the main text. Moreover, from the output of the numerics, we could in certain cases reconstruct analytically the trilocal models. Below we detail some of these models. Notably, these models can saturate some of the NSI constraints that we have derived, implying that the latter are tight. Finally, we also show that any point within these trilocal regions can be achieved via a trilocal model. In other words, although we characterize only the boundary of these regions, we show that any point within the boundary is also achievable (i.e.~these regions do not feature any hole).

We first discuss a simple class of trilocal models, featuring only binary shared variables:
\begin{align}
	P = \left[
	\begin{array}{c|cc}
		& 1 & 2 \\
		\hline
		\alpha & r & 1-r \\
		\beta & q & 1-q \\
		\gamma & p & 1-p
	\end{array} \right], \;
	f_a = f_b = f_c =
	\begin{bmatrix}
		1 & 0 \\
		0 & 0
	\end{bmatrix}.
\end{align}
This results in the following statistics. The single-party marginals are given by
\be
E_\mathrm{A} = 2pq-1 \quad E_\mathrm{B} = 2pr-1 \quad E_\mathrm{C} = 2qr-1 \,.
\ee
Next, the two-body marginals are
\ba
E_\mathrm{AB} = 1- 2pq -2pr +4pqr  \\
E_\mathrm{BC} = 1- 2rp -2rq +4pqr  \\
E_\mathrm{AC} = 1- 2qp -2qr +4pqr \,
\ea
while the three-body correlator is
\be
E_\mathrm{ABC} = -1 + 2pq +2pr +2qr -4pqr \,.
\ee
This model can saturate Supplementary Inequality \eqref{SM:nice2}, for the case where single-party and two-body marginals are fully symmetrical. This shows that \eqref{SM:nice2} represents a tight constraint for NSI. Here, we take simply $p=q=r $ (i.e.~all sources are equivalent). This leads to
\ba
E_1 &=& 2 p^2 -1 \\
E_2 &=& 1-4p^2 +4p^3
\ea
which corresponds to the equality condition in Eq.~\eqref{SM:nice2} (assuming here $p \geq 1/\sqrt{2}$, so that $E_1 \geq 0$). Note that for $p=1/\sqrt{2}$, we get $E_1=0$, $E_2 = \sqrt{2}-1$ and $E_3= 2  -\sqrt{2}$. This model corresponds to the top-right point of the white region in Fig.~3 of the main text.

Note also that this class of trilocal models (for arbitrary values of $p$, $q$ and $r$) saturate the conjectured NSI constraint of Eq.~(8) of the main text.

Next, we present a trilocal model that achieves the $E_2=-\frac{1}{3}$ and $E_1=0$:
\begin{align}
	P &= \left[
	\begin{array}{c|cc}
		& 1 & 2 \\
		\hline
		\alpha & 1/3 & 2/3 \\
		\beta & 3/4 & 1/4 \\
		\gamma & 2/3 & 1/3
	\end{array} \right],  \; \nonumber \\
	f_a &=
	\begin{bmatrix}
		1 & 0 \\
		0 & 0
	\end{bmatrix}, \;
	f_b =
	\begin{bmatrix}
		0 & 1/2 \\
		1/2 & 1
	\end{bmatrix},\;
	f_c =
	\begin{bmatrix}
		1 & 1 \\
		0 & 1
	\end{bmatrix}.
\end{align}

Moreover, from Fig.~2 of the main text, we see that there exist trilocal models with $E_2=-\frac{1}{3}$ and strictly positive $E_1$. We find that the model that maximizes $E_1$ (while keeping $E_2=-\frac{1}{3}$) is given by:
\begin{align}
	P & = \left[
	\begin{array}{c|ccc}
		& 1 & 2 & 3\\
		\hline
		\alpha & x & 1-x & 0\\
		\beta & y & (1-y)/2 & (1-y)/2\\
		\gamma & 1-x & x & 0
	\end{array} \right], \; \nonumber \\
	f_a & =
	\begin{bmatrix}
		1 & 0 & 1\\
		0 & 0 & 0\\
		1 & 1 & 1\\
	\end{bmatrix}, \;
	f_b =
	\begin{bmatrix}
		1 & 1 & 0 \\
		0 & 1 & 0\\
		0 & 0 & 0
	\end{bmatrix},\;
	f_c =
	\begin{bmatrix}
		0 & 1 & 0 \\
		1 & 1 & 0\\
		0 & 0 & 0
	\end{bmatrix},
\end{align}
where $x$ is the root between $0$ and $1$ for $3x^4-9x^3+9x^2-5x+1=0$, and $y=\frac{1}{3(2x^2-2x+1)}$. Consequently, $E_1 = (3y^3+y^2+y-1)/4 \approx 0.1753$.

\jd{Related to Fig.~3 of the main text, we now} give a trilocal model for $E_1=E_3=0$ and $E_2\approx 0.3621$:
\begin{align}
	P & = \left[
	\begin{array}{c|ccc}
		& 1 & 2 & 3\\
		\hline
		\alpha,\beta,\gamma & x & y & z
	\end{array} \right], \; \nonumber \\
	f_a & = f_b = f_c =
	\begin{bmatrix}
		0 & 0 & 1\\
		0 & 0 & 0\\
		1 & 0 & 1\\
	\end{bmatrix},
\end{align}
where $x=\frac23 - \frac{z^3}{3} - \frac{z}{2}$, $y=\frac13 + \frac{z^3}{3} - \frac{z}{2}$, and $z\approx 0.3861$ is the root between $0$ and $1$ for $-4z^7+4z^4-3z^3+8z-3=0$. Consequently, $E_2 = \frac49 z^7 - \frac83 z^5 + \frac89 z^4 - z^3 + \frac{16}{3} z^2 - \frac{32}{9}z + 1 \approx 0.3621$. Note that by changing the distributions of the shared variables (but keeping the local response functions the same), one can generate the correlations indicated by the blue dots in Fig.~3 of the main text (for $0<E_3 \leq 2-\sqrt{2}$).

Finally, we now show that the trilocal regions shown in the various figures of the paper do not feature any hole. That is, we have characterized the boundary of these trilocal regions (in some cases by giving explicit trilocal models), and we now prove that any point inside this boundary can necessarily be achieved by a trilocal model.

The idea is to consider the following ``depolarizing'' protocol. Consider a distribution $p_0(abc)$ achievable via a trilocal model $M$, given by single-party marginals $E_\mathrm{A}^0$, etc..., bipartite marginals $E_\mathrm{AB}^0$, etc...,  and a tripartite correlator $E_\mathrm{ABC}^0$. Each party adds noise locally (and independently of the other parties) via the following procedure. With probability $1-\eta$ a party provides a random output, while with probability $\eta$ they output according to $M$. Hence we obtain the continuous family of distributions characterized by $E_\mathrm{A}= \eta E_\mathrm{A}^0$ etc, $E_\mathrm{AB}= \eta^2 E_\mathrm{AB}^0$ etc and $E_\mathrm{ABC}= \eta^3 E_\mathrm{ABC}^0$. Varying $\eta$ from 1 to 0 we obtain a continuous curve from $p_0(abc)$ to the uniform distribution.

\begin{figure}[t!]
\includegraphics[width = \columnwidth]{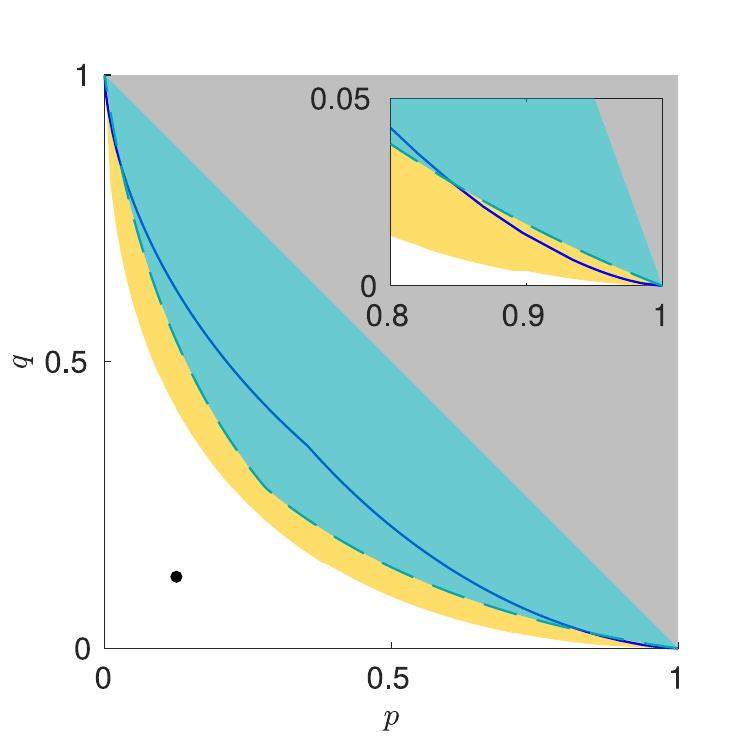}
\caption{Comparison of our NSI constraint \eqref{SM:nice2} and the Finner inequality \eqref{Finner} of Ref.~\cite{Renou} (solid blue curve represents equality in \eqref{Finner}), for the set of distributions $p_{p,q}$ given in Eq.~\eqref{pq}. Our NSI constraints (dashed turquoise curve representing equality in \eqref{SM:nice2}) appears to be stronger almost everywhere, except for two small regions, around each deterministic point $P_{+++}$ ($p=1$) and $P_{---}$ ($q=1$); see inset. As in previous figures, the grey region is excluded via positivity constraints and the turquoise region via NSI constraints. The white region is achievable via trilocal models, while the yellow region is undetermined. The black point represents the uniformly random distribution, i.e.~$p=q=1/8$.}
\label{finner_fig}
\end{figure}

For Figs. 2 and 3 of the main text and Supplementary Figure~\ref{fig:EabvsEbc}, we see that any point inside the trilocal region can be obtained by adding (a well chosen) amount of noise to a distribution sitting on the boundary. For Supplementary Figure~\ref{finner_fig}, the situation is different, as the above depolarizing procedure takes an initial distribution on the slice outside of it.

\section*{Appendix 3: Comparison to Finner inequality}

We compare the NSI constraint derived here with a criterion derived in Ref.~\cite{Renou}. The later is based on the Finner inequality, and states that
\be \label{Finner}
p(abc) \leq \sqrt{ p_A(a)p_B(b)p_C(c)}
\ee
where $p_A(a)$ represents Alice's marginal probability to observe outcome $a$, and similarly for $p_B(b)$ and $p_C(c)$. Importantly, it should be pointed out that the above inequality is only conjectured to hold under NSI. At the moment, it is only proven that inequality \eqref{Finner} holds in quantum theory, and in ``boxworld'' (a generalized probabilistic theory where sources can prepare arbitrary no-signalling boxes, see e.g.~\cite{barrett}). This does not imply that \eqref{Finner} holds in any generalized probabilistic theory.

Nevertheless it is interesting to compare both approaches, in particular as inequality \eqref{Finner} involves explicitly tripartite correlations (e.g.~the term $p(abc)$), contrary to our approach which seems to be limited to single- and two-party marginals.

To perform this comparison, we use a specific set of distributions in the triangle network, discussed in \cite{Renou}. These take the form
\begin{equation}\label{pq}
p_{p,q} = p P_{+++} + q P_{---} + (1-p-q) P_{\text{diff}}
\end{equation}
where $P_{abc}$ represents the distribution where the outputs are set to values $a$, $b$ and $c$ deterministically, and $P_{\text{diff}} = (P_{++-} + P_{+-+}+ P_{-++}+ P_{+--}+ P_{-+-}+ P_{--+})/6$. From \eqref{Finner} it follows that $p_{p,q} $ is not realizable in the triangle network when $q> 1+p -2p^{2/3}$ (and a similar constraint inverting $p$ and $q$); given by the black curve in Supplementary Figure~\ref{finner_fig}.

We compare this criterion to our NSI constraint \eqref{SM:nice2}. From Supplementary Figure~\ref{finner_fig}, we see that our NSI constraint is mostly stronger than the Finner inequality \eqref{Finner}. However, this is not the case in general, as there is a small region (around the deterministic points $P_{+++}$ and $P_{---}$) where the Finner inequality is stronger.

\section*{Appendix 4: Proof of a general NSI inequality}
\jd{
Here we prove the validity of the following inequality for NSI models in the triangle network:
\begin{equation}\label{eq:conjecture}
\begin{split}
&(1+|E_\mathrm{A}|+|E_\mathrm{B}|+E_\mathrm{AB})^2 \\
&+ (1+|E_\mathrm{A}|+|E_\mathrm{C}|+E_\mathrm{AC})^2  \\
&+ (1+|E_\mathrm{B}|+|E_\mathrm{C}|+E_\mathrm{BC})^2 \\
&\leq 6(1+|E_\mathrm{A}|)(1+|E_\mathrm{B}|)(1+|E_\mathrm{C}|) \,.
\end{split}
\end{equation}

This inequality is invariant under exchange of parties. It is also invariant under the joint relabelling of all parties' outputs. However, it is not invariant under arbitrary output relabelling. Therefore, we consider two cases:
\begin{enumerate}
\item If $E_\mathrm{A}, E_\mathrm{B}, E_\mathrm{C} \geq 0$, Supplementary Equation~\eqref{eq:conjecture} can be written
\begin{equation}\label{eq:conjecture1}
\begin{split}
&(1+E_\mathrm{A}+E_\mathrm{B}+E_\mathrm{AB})^2 \\
&+ (1+E_\mathrm{A}+E_\mathrm{C}+E_\mathrm{AC})^2  \\
&+ (1+E_\mathrm{B}+E_\mathrm{C}+E_\mathrm{BC})^2 \\
&\leq 6(1+E_\mathrm{A})(1+E_\mathrm{B})(1+E_\mathrm{C}) \,.
\end{split}
\end{equation}
If $E_\mathrm{A}, E_\mathrm{B}, E_\mathrm{C} \leq 0$, flipping all outcomes brings us back to the same condition.
\item In all other cases we can always exchange parties and outcomes (jointly) to reach the case $E_\mathrm{A}, E_\mathrm{C} \geq 0$, $E_\mathrm{B} \leq 0$. In this case, Supplementary Equation~\eqref{eq:conjecture} reduces to
\begin{equation}\label{eq:conjecture2}
\begin{split}
&(1+E_\mathrm{A}-E_\mathrm{B}+E_\mathrm{AB})^2 \\
&+ (1+E_\mathrm{A}+E_\mathrm{C}+E_\mathrm{AC})^2  \\
&+ (1-E_\mathrm{B}+E_\mathrm{C}+E_\mathrm{BC})^2 \\
&\leq 6(1+E_\mathrm{A})(1-E_\mathrm{B})(1+E_\mathrm{C}) \,.
\end{split}
\end{equation}
\end{enumerate}
It is thus sufficient to show the validity of the two inequalities~\eqref{eq:conjecture1} and~\eqref{eq:conjecture2} in their respective context.

Before focusing on these cases, we make some general observations. Since probabilities are positive, the following sum of probabilities also is:
\begin{equation}
\begin{split}
&p (-1,1,-1,1,1,1) + p(1,-1,1,-1,-1,-1)\\
&+ p(-1,-1,1,-1,1,1) + p(1,1,-1,1,-1,-1)\geq 0.\\
\end{split}
\end{equation}
Using Supplementary Equation~\eqref{eq:hexagonDecomp}, this condition can be rewritten as the following inequality:
\begin{equation}\label{eq:one}
  1 - 2 E_\mathrm{AC} + 2 E_\mathrm{B} F_3  \geq  E_\mathrm{AB}^2 -  E_\mathrm{AC}^2 + E_\mathrm{BC} ^2.
\end{equation}
Similarly, $\sum_{a',b',c'} p(1,-1,1,a',b',c') \geq 0$
implies
\begin{equation}\label{eq:I_two}
 I_2 = 1 + E_\mathrm{A} - E_\mathrm{B} + E_\mathrm{C} + E_\mathrm{A} E_\mathrm{C} - E_\mathrm{AB} - E_\mathrm{BC} - F_3 \geq 0,
\end{equation}
and $\sum_{a',b',c'} p(-1,1,-1,a',b',c') \geq 0$
implies
\begin{equation}\label{eq:I_twop}
 I_2' = 1 - E_\mathrm{A} + E_\mathrm{B} - E_\mathrm{C} + E_\mathrm{A} E_\mathrm{C} - E_\mathrm{AB} - E_\mathrm{BC} + F_3 \geq 0,
\end{equation}

\bigskip

\textit{Case 1 ($E_\mathrm{A}, E_\mathrm{B}, E_\mathrm{C} \geq 0$):} Since $E_\mathrm{B} \geq 0$, the product $E_\mathrm{B} I_2$ \eqref{eq:I_two} is positive. After rearrangement, we obtain
\begin{equation}
\begin{split}
& 2 (1 + E_\mathrm{A})(1+ E_\mathrm{B})(1 + E_\mathrm{C}) - (1+E_\mathrm{A}+E_\mathrm{C}+E_\mathrm{AC})^2\\
& \geq  -E_\mathrm{A}^2 +2 E_\mathrm{B}^2 -E_\mathrm{C}^2 + 2 E_\mathrm{B} (E_\mathrm{AB} + E_\mathrm{BC}) \\
& \ \ \ - 2 E_\mathrm{AC} (E_\mathrm{A} +E_\mathrm{C}) - E_\mathrm{AC}^2 + 1 - 2E_\mathrm{AC} + 2E_\mathrm{B} F_3.
\end{split}
\end{equation}
Recognizing the last three terms from~\eqref{eq:one}, we use this inequality to eliminate the term containing the free variable $F_3$ and get
\begin{equation}\label{eq:var1}
\begin{split}
& 2 (1 + E_\mathrm{A}) (1 + E_\mathrm{B}) (1+E_\mathrm{C}) - (1 + E_\mathrm{A} + E_\mathrm{C} + E_\mathrm{AC})^2 \\
& \geq - E_\mathrm{A}^2 + 2 E_\mathrm{B}^2 - E_\mathrm{C}^2 + 2 E_\mathrm{B} (E_\mathrm{AB}+E_\mathrm{BC}) \\
& \ \ \ - 2 E_\mathrm{AC} (E_\mathrm{A}+E_\mathrm{C}) + E_\mathrm{AB}^2 - 2 E_\mathrm{AC}^2 + E_\mathrm{BC}^2.
\end{split}
\end{equation}

Since $E_\mathrm{C}$ is positive, this inequality remains valid if we cyclically permute the parties to let $C$ play the role of $B$, yielding
\begin{equation}\label{eq:var2}
\begin{split}
& 2 (1 + E_\mathrm{A}) (1 + E_\mathrm{B}) (1+E_\mathrm{C}) - (1 + E_\mathrm{A} + E_\mathrm{B} + E_\mathrm{AB})^2 \\
& \geq - E_\mathrm{A}^2 -  E_\mathrm{B}^2 + 2 E_\mathrm{C}^2 + 2 E_\mathrm{C} (E_\mathrm{AC}+E_\mathrm{BC}) \\
& \ \ \ - 2 E_\mathrm{AB} (E_\mathrm{A}+E_\mathrm{B})  - 2 E_\mathrm{AB}^2 + E_\mathrm{AC}^2 + E_\mathrm{BC}^2.
\end{split}
\end{equation}
Similarly, since $E_\mathrm{A}\geq 0$ we can also write
\begin{equation}\label{eq:var3}
\begin{split}
& 2 (1 + E_\mathrm{A}) (1 + E_\mathrm{B}) (1+E_\mathrm{C}) - (1 + E_\mathrm{B} + E_\mathrm{C} + E_\mathrm{BC})^2 \\
& \geq 2 E_\mathrm{A}^2 - E_\mathrm{B}^2 - E_\mathrm{C}^2 + 2 E_\mathrm{A} (E_\mathrm{AB}+E_\mathrm{AC}) \\
& \ \ \ - 2 E_\mathrm{BC} (E_\mathrm{B}+E_\mathrm{C}) + E_\mathrm{AB}^2 + E_\mathrm{AC}^2 - 2 E_\mathrm{BC}^2.
\end{split}
\end{equation}
Summing up the three Supplementary Equations~\eqref{eq:var1}, \eqref{eq:var2}, \eqref{eq:var3}, all terms on the right-hand side cancel out, leaving us with:
\begin{equation}
\begin{split}
& 6 (1 + E_\mathrm{A}) (1 + E_\mathrm{B}) (1+E_\mathrm{C}) - (1 + E_\mathrm{A} + E_\mathrm{C} + E_\mathrm{AC})^2 \\
& - (1 + E_\mathrm{A} + E_\mathrm{B} + E_\mathrm{AB})^2 - (1 + E_\mathrm{B} + E_\mathrm{C} + E_\mathrm{BC})^2 \geq 0.
\end{split}
\end{equation}
This inequality is identical to~\eqref{eq:conjecture1}, which concludes the proof under the $E_\mathrm{A} \geq 0, E_\mathrm{B} \geq 0, E_\mathrm{C} \geq 0$ assumption.

\bigskip

\textit{Case 2 ($E_\mathrm{A}, E_\mathrm{C} \geq 0$, $E_\mathrm{B}\leq 0$):} This time, $E_\mathrm{B}$ is negative, so we cannot use Supplementary Equation~\eqref{eq:var1}. However, we still have $E_\mathrm{A} \geq 0$ and $E_\mathrm{C} \geq 0$ so Supplementary Equations~\eqref{eq:var2} and \eqref{eq:var3} remain valid. For clarity, we rearrange them as
\begin{equation}\label{eq:var2bis}
\begin{split}
&2(1+E_\mathrm{A})(1-E_\mathrm{B})(1+E_\mathrm{C}) - (1+E_\mathrm{A}-E_\mathrm{B}+E_\mathrm{AB})^2\\
&\geq -E_\mathrm{A}^2 - E_\mathrm{B}^2 + 2E_\mathrm{C}^2 - 2E_\mathrm{AB}^2 + E_\mathrm{BC}^2 + E_\mathrm{AC}^2\\
&\ \ \ + 2 E_\mathrm{B} E_\mathrm{AB} - 2 E_\mathrm{A} E_\mathrm{AB} + 2 E_\mathrm{C} (E_\mathrm{AC} + E_\mathrm{BC})\\
&\ \ \ - 4 E_\mathrm{B} (E_\mathrm{C} + E_\mathrm{A} E_\mathrm{C})
\end{split}
\end{equation}
and
\begin{equation}\label{eq:var3bis}
\begin{split}
&2(1+E_\mathrm{A})(1-E_\mathrm{B})(1+E_\mathrm{C}) - (1-E_\mathrm{B}+E_\mathrm{C}+E_\mathrm{BC})^2\\
&\geq 2 E_\mathrm{A}^2-E_\mathrm{B}^2-E_\mathrm{C}^2 + E_\mathrm{AB}^2-2 E_\mathrm{BC}^2+E_\mathrm{AC}^2\\
&\ \ \  + 2 E_\mathrm{B} E_\mathrm{BC} -2 E_\mathrm{C} E_\mathrm{BC} + 2 E_\mathrm{A} (E_\mathrm{AB}+E_\mathrm{AC})\\
&\ \ \ - 4 E_\mathrm{B} (E_\mathrm{A}+E_\mathrm{A}E_\mathrm{C})
\end{split}
\end{equation}
Considering now the positive expression $I_2'$ in Supplementary Equation~\eqref{eq:I_twop}, we multiply it with $-E_\mathrm{B}$, which is positive. After rearrangement, this gives
\begin{equation}
\begin{split}
&2(1+E_\mathrm{A})(1-E_\mathrm{B})(1+E_\mathrm{C}) - (1+E_\mathrm{A}+E_\mathrm{C}+E_\mathrm{AC})^2\\
&\geq - E_\mathrm{A}^2 + 2E_\mathrm{B}^2 - E_\mathrm{C}^2 - 2 E_\mathrm{B} (E_\mathrm{AB}+E_\mathrm{BC})\\
&\ \ \  - 2 E_\mathrm{AC} (E_\mathrm{A}+E_\mathrm{C})  - 4 E_\mathrm{B} (E_\mathrm{A}+E_\mathrm{C}) - E_\mathrm{AC}^2\\
&\ \ \  + 1-2 E_\mathrm{AC} + 2 E_\mathrm{B} F_3
\end{split}
\end{equation}
Recognizing again the last three terms from Supplementary Equation~\eqref{eq:one}, we obtain
\begin{equation}\label{eq:lasteq}
\begin{split}
&2(1+E_\mathrm{A})(1-E_\mathrm{B})(1+E_\mathrm{C}) - (1+E_\mathrm{A}+E_\mathrm{C}+E_\mathrm{AC})^2\\
&\geq -E_\mathrm{A}^2 + 2 E_\mathrm{B}^2 - E_\mathrm{C}^2 - 2E_\mathrm{B} (E_\mathrm{AB}+E_\mathrm{BC})\\
&\ \ \  - 2E_\mathrm{AC}(E_\mathrm{A}+E_\mathrm{C})  - 4E_\mathrm{B}(E_\mathrm{A}+E_\mathrm{C})\\
&\ \ \ + E_\mathrm{AB}^2 + E_\mathrm{BC}^2 - 2E_\mathrm{AC}^2
\end{split}
\end{equation}

Summing Supplementary Equations~\eqref{eq:var2bis},~\eqref{eq:var3bis} and~\eqref{eq:lasteq}, we obtain
\begin{equation}
\begin{split}
& 6 (1 + E_\mathrm{A}) (1 - E_\mathrm{B}) (1+E_\mathrm{C}) - (1 + E_\mathrm{A} + E_\mathrm{C} + E_\mathrm{AC})^2 \\
& - (1 + E_\mathrm{A} - E_\mathrm{B} + E_\mathrm{AB})^2 - (1 - E_\mathrm{B} + E_\mathrm{C} + E_\mathrm{BC})^2 \\
& \geq -8 E_\mathrm{B} (E_\mathrm{A}+E_\mathrm{C}+E_\mathrm{A} E_\mathrm{C}) \geq 0,
\end{split}
\end{equation}
where the positivity follows from the negativity of $E_\mathrm{B}$ and the positivity of $E_\mathrm{A}$ and $E_\mathrm{C}$. This proves~\eqref{eq:conjecture2} under the  $E_\mathrm{A} \geq 0, E_\mathrm{B} \leq 0, E_\mathrm{C} \geq 0$ assumptions, which concludes the proof of Supplementary Equation~\eqref{eq:conjecture}.
}

\section*{References}

\bibliographystyle{naturemag}
\bibliography{Triangle5}

\end{document}